# Intrinsic Lorentz violation in Doppler effect from a moving point light source

Changbiao Wang (changbiao_wang@yahoo.com)
*ShangGang Group, 70 Huntington Road, Apartment 11, New Haven, CT 06512, USA*

**Abstract.** — Einstein's Doppler formula is not applicable when a moving point light source is close enough to the observer; for example, it may break down or cannot specify a determinate value when the point source and the observer overlap. In this paper, Doppler effect for a moving point light source is analyzed, and it is found that the principle of relativity allows the existence of intrinsic Lorentz violation. A conceptual scheme to experimentally test the point-source Doppler effect is proposed, and such a test could lead to an unexpected result that the frequency of a photon may change during propagation, which questions the constancy of Planck constant since the energy conservation in Einstein's light-quantum hypothesis must hold.

## I. INTRODUCTION

Principle of relativity and constancy of the light speed in free space are the two basic postulates of the special theory of relativity [1,2]. A uniform plane electromagnetic wave, which is a fundamental solution to Maxwell equations, propagates at the light speed in all directions [3]. Consequently, when directly applying the relativity principle to Maxwell equations, one may find that the light speed must be the same in all inertial frames of reference, in other words, the covariance of Maxwell equations requires the constancy of light speed [4].

According to the principle of relativity, the phase factor $\exp i(\omega t - \mathbf{k} \cdot \mathbf{x})$ of a plane wave in free space is symmetric with respect to all inertial frames, where $t$ is the time, $\mathbf{x}$ is the position vector in space, $\omega$ is the frequency, and $|\mathbf{k}| = \omega/c$ is the wave number with $c$ the universal light speed. $(\mathbf{x}, ct)$ and $(\mathbf{k}, \omega/c)$ are independent. Since the phase $\Psi = (\omega t - \mathbf{k} \cdot \mathbf{x})$ is a Lorentz invariant and $X^\mu = (\mathbf{x}, ct)$ is a Lorentz covariant four-vector, $K^\mu = (\mathbf{k}, \omega/c)$ must be covariant, and the phase function can be written in a standard covariant form, given by $\Psi = g_{\mu\nu} K^\mu X^\nu = (\omega t - \mathbf{k} \cdot \mathbf{x})$, with the metric tensor $g_{\mu\nu} = g^{\mu\nu} = diag(-1,-1,-1,+1)$ [5]. Thus the Doppler formula for a plane wave can be directly obtained from the Lorentz transformation of $K^\mu = (\mathbf{k}, \omega/c)$ [6].

For a spherical wave in free space, which is generated from a moving point light source, such as a radiation electric dipole [3,7-10], the phase function is given by $\Psi = \omega t - |\mathbf{k}||\mathbf{x}|$ in the source-rest frame, and it is also a Lorentz invariant. But there is an additional strong constraint between $(\mathbf{x}, ct)$ and $(\mathbf{k}, \omega/c)$; $\mathbf{k} \cdot \mathbf{x} = |\mathbf{k}||\mathbf{x}|$ must hold. As a result, the Lorentz covariance of $(\mathbf{k}, \omega/c)$ cannot hold any more (see Appendix A). Thus the Doppler formula for a moving point light source cannot be obtained from the Lorentz transformation of $(\mathbf{k}, \omega/c)$. In other words, $(\mathbf{k}, \omega/c)$ does not follow Lorentz transformation; this physical phenomenon is termed to be "intrinsic Lorentz violation" (or "intrinsic breaking of Lorentz invariance") in this paper.

Obviously, Einstein's plane-wave Doppler formula is not applicable when a moving point light source is close enough to the observer; for example, it may break down or cannot specify a determinate value when the point source and the observer overlap (confer Appendix B).

At first thought, one might question "the overlap of a point source with the observer" as being a really absurd statement, and also question the validity of the spherical-wave model when the observer is in the near-field zone. On second thoughts, one may find that those challenges actually put the validity of Lorenz transformation into question. As we remember, it is the point light source that Einstein used to derive the time-space Lorentz transformation: When $t' = t = 0$ and $\mathbf{x}' = \mathbf{x} = 0$, a spherical wave is fired …[1]; obviously, the two observers and the point source are overlapped at $t' = t = 0$.

It is well known from the classical electromagnetic theory [3,7-10] that, the spherical form of wavefronts from the electric dipole radiation is valid at any distances. In the far-field zone, the radiation field ($\sim 1/|\mathbf{x}|$) is dominant in strength, while in the near-field zone, the quasi-Biot-Savart induction field ($\sim 1/|\mathbf{x}|^2$) and the dipole quasi-Coulomb field ($\sim 1/|\mathbf{x}|^3$) are dominant. The far-field and the near-fields cannot exist independently and they are together as a whole to satisfy with-source Maxwell equations, so that all the fields (waves) have the same frequency (wavelength).

Nevertheless, one might still insist that Einstein's plane wave Doppler formula be applicable to any cases, no matter whether the observer is close to a moving point source or not. A strong argument is that any spherical wave can be decomposed into plane waves. Unfortunately, however, that is not true. A point-source-generated (traveling) spherical wave in free space, like the plane wave, is monochromatic and non-dispersive, with group velocity equal to phase velocity, and it *cannot* be decomposed into a combination of *physical* plane waves. Moreover, even if the spherical wave could be decomposed into plane waves, we still could not obtain the Doppler frequency shift of the spherical wave from component plane waves, because all the component plane waves propagate in *different directions* (otherwise not a spherical wave), while Doppler effects depend on individual plane-wave propagation directions, and how to define the frequency of the whole spherical wave becomes questionable. Therefore, the plane-wave decomposition is not viable in solving this problem.

Fundamental relativistic time-space consequences such as the relativity of simultaneity, time dilation, Lorentz contraction, and Doppler frequency shift for a plane wave can be derived by making use of Lorentz transformation [1], a standard analytical approach. As mentioned above, however, the point-source Doppler formula cannot be obtained from the Lorentz transformation. Thus a "direct approach" without using the Lorentz transformation becomes indispensable.

In this paper, to better understand profound implications of Einstein's relativity, Doppler formula for a moving point light source is derived with a direct approach. A conceptual experimental scheme to test the formula is proposed.

An important significance of the point-source Doppler effect is that it predicts a new physics: intrinsic Lorentz violation.

The paper is organized as follows. In Sec. II, by introduction of the *invariance of event number*, a spherical-mirror light clock is used to re-examine all the relativity of simultaneity, time dilation, and Lorentz contraction in the *same* thought experiment. The purpose is to show how to catch the time dilation effect in the direct approach. In Sec. III, the Doppler formula for a moving point light source is developed, and it is used to analyze previously-published experimental results. In Sec. IV, conclusions and remarks are given, and the traditional understanding of the principle of relativity is reviewed. In Appendix A, it is shown why the Lorenz covariance of $(\mathbf{k}, \omega/c)$ for a moving point light source is violated; in Appendix B, an unconventional "short-range" longitudinal Doppler effect is shown; in Appendix C, a conceptual experimental scheme for verifying the point-source Doppler effect is presented.

## II. A SPHERICAL LIGHT-CLOCK THOUGHT EXPERIMENT

In this section, a thought experiment, in which a light clock has a spherical mirror with a proper radius of $R_0$ (see Fig. 1), is presented to show the relativity of simultaneity, time dilation, and Lorentz contraction. The purpose is to help understand the time dilation effect in the "direct approach" for deriving relativistic results where Lorentz transformations may not apply.

Suppose that a flash of light is emitted at the center $O'$ of the mirror. All the rays in different directions reach different locations of the mirror surface at the same time, observed by the $O'$-observer, and they are returned to the center also at the same time. The emitting (receiving) is counted as one event; namely, it is one event for all the rays to start (end) at the same place and the same time. According to the relativity principle, *the event number must be invariant*; consequently, observed in *any* inertial frames, all the rays generated by the above flash start (end) at the same place and the same time.

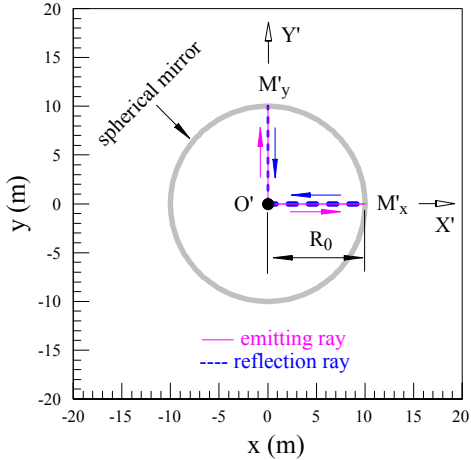

Fig. 1. Spherical-mirror light clock (cross section) at rest, which has a spherical mirror with a radius of $R_0$. A flash of light is emitted at the center $O'$ and returned after a time of $\Delta t' = 2R_0/c$, observed by the $O'$-observer. The emitting and reflection rays in all directions have an identical length of $R_0$. $O'M_y'$- and $M_y'O'$-rays are used to determine time dilation; $O'M_x'$- and $M_x'O'$-rays are used to determine Lorentz contraction.

Suppose that the spherical-mirror light clock moves relatively to the $O$-observer in the lab frame at a uniform velocity of $v = \beta c$ with $c$ the light speed. When $O'$ overlaps $O$, the $O'$-observer emits a flash and receives it after a proper time interval of $\Delta t' = 2R_0/c$, observed by the $O'$-observer, and all the rays leave $O'$ and they are returned to $O'$, respectively at the same times. According to the invariance of event number, observed by the $O$-observer, all the rays start at $O$ and end at $O'$, also respectively at the same times, with a time interval of $\Delta t$; the two events take place at different places, separated by a distance of $OO' = v\Delta t$. Thus all the rays in different directions, reflected by the mirror, go an identical total distance of $c\Delta t$ according to the constancy of light speed. From analytical geometry, the set of points whose distances from the two points $O$ and $O'$ have a constant sum of $c\Delta t$ is a prolate ellipsoid of revolution, as shown in Fig. 2. This prolate ellipsoid is a collection of all the points at which the mirror reflects the emitting rays at different times, while the moving mirror, measured by the $O$-observer at the same time, is an oblate ellipsoid of revolution.

Since the length perpendicular to the direction of motion is assumed to be the same [1], the major and minor axes of the prolate ellipsoid are, respectively, $c\Delta t/2$ and $R_0$ long. From Fig.1 and Fig. 2, we can see that, observed by the $O'$-observer, all the emitting rays reach the mirror surface at the same time, while observed by the $O$-observer, all the emitting rays have different lengths and they reach the mirror surface in different times. Thus the relativity of simultaneity is clearly shown.

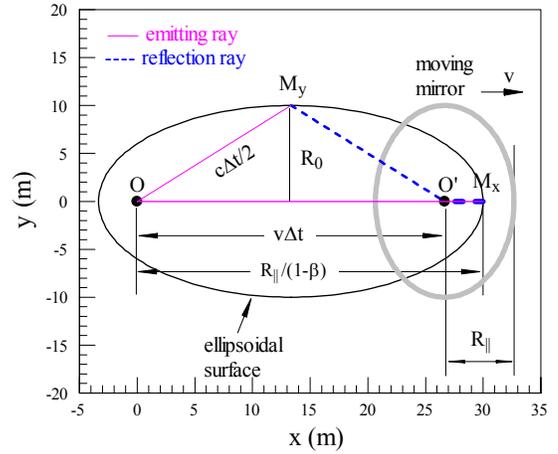

Fig. 2. Spherical-mirror light clock (cross section) in motion, at a velocity of $v$ relatively to the $O$-observer. When $O'$ overlaps $O$, the $O'$-observer emits a flash and receives the flash reflected by the mirror after a time of $\Delta t$, observed by the $O$-observer. Emitting rays have different lengths and reach a prolate ellipsoidal surface at different times. The moving mirror is compressed in the direction of motion into Einstein's oblate ellipsoid of revolution [1]. The figure was drawn with $R_0 = 10$ m and $\beta = 0.8$.

$O'M_y'$ and $M_y'O'$ in Fig. 1 correspond to $OM_y$ and $M_yO'$ in Fig. 2, which is exactly the same as the plane-plate light-clock case [11], and we obtain the time dilation expression, given by $\Delta t = \gamma(2R_0/c) = \gamma \Delta t'$, with $\gamma = (1-\beta^2)^{-1/2}$ the time-dilation factor.

$O'M_x'$ and $M_x'O'$ in Fig. 1 correspond to $OM_x$ and $M_xO'$ in Fig. 2. Suppose that the time intervals, required by the light flash to go from $O$ to $M_x$ and from $M_x$ to $O'$, are $\delta t_1$ and $\delta t_2$ respectively, and the mirror radius in the direction of motion is $R_\parallel$. Following the way suggested by Kard [12] to calculate the distance a light signal goes over a moving rod, we have $OM_x = c\delta t_1 = R_\parallel + v\delta t_1$ and $M_xO' = c\delta t_2 = R_\parallel - v\delta t_2$, leading to $OM_x = R_\parallel/(1-\beta)$ and $M_xO' = R_\parallel/(1+\beta)$. Since $\delta t_1 + \delta t_2 = \Delta t = \gamma(2R_0/c)$ and $OM_x + M_xO' = c\Delta t$, we obtain the Lorentz contraction expression, given by $R_\parallel = R_0/\gamma$.

From the above thought experiment we can see that the time interval of two events occurring at the same place is the shortest,

namely a time-dilation effect ($\Delta t = \gamma \Delta t'$) [1]. Since the thought experiment is applicable to any observers of relative inertial motion, the time-dilation effect holds for any two of the events occurring at the same place. Compared with the Lorentz contraction, the time dilation has a more straightforward definition, and it is a core result of the relativity principle. When a direct approach is used to derive relativistic results, grasping the time-dilation effect is a key point, which can be seen in the following derivation of Doppler formula.

### III. RELATIVISTIC DOPPLER EFFECT FOR A MOVING POINT LIGHT SOURCE

Einstein derived Lorentz transformation by use of a spherical wave and developed Doppler formula for a plane wave [1]. As we have known, the light speed $c$ has no preferred frame, no matter for a plane wave or a spherical wave. But the moving point source has a preferred frame, in which all the spherical wavefronts take the point source as a common center. Because of this, the Doppler formula for a moving point light source is different from the one for a plane wave.

The observed wave period $T$ is defined as the time interval of two consecutive wave-crests that the observer receives at the *same* place, the frequency is defined as $\omega = 2\pi/T$, and the wavelength is defined as $\lambda = cT$; this definition is a generalization of the one for a plane wave [4]. However, it should be pointed out that, for a plane wave, observed in any given inertial frame, the wave vector and frequency are the same everywhere, while for a moving point source, observed in a frame moving relatively to the point source, the wave vector and frequency *depend on the location and time*.

In above, we use "two consecutive wave-crests" to describe the definition; actually it should be understood as "two consecutive phases with a phase difference of $2\pi$".

The Doppler effect of wave period actually describes the relation between *the time interval* in which the moving observer emits two consecutive $\delta$-light signals and *the time interval* in which the lab observer receives the two $\delta$-signals at the same place. Accordingly, the wave period is a measurable quantity everywhere in principle.

Suppose that a point light source fixed in $X'O'Y'$ frame moves relatively to the observer fixed in $XOY$ frame, as shown in Fig. 3. Observed in the $XOY$ frame, the light source generates two consecutive spherical crest-wavefronts at the times $t = t_1$ and $t_2$ respectively, with a separation of $O'_1O'_2 = (t_2 - t_1)|\boldsymbol{\beta}|c$. The observer receives the two consecutive crest-signals at the different retarded times $t_{1r} = t_1 + R_1/c$ and $t_{2r} = t_2 + R_2/c$ at the *same* place, and the observed wave period is given by $T = t_{2r} - t_{1r}$. Observed in the light-source $X'O'Y'$ frame, the time interval of the two consecutive crest-wavefronts, which are generated in the *same* place, is the wave period, given by $T' = t'_2 - t'_1$. As shown in Sec. II, between two observers of relative motion, there is a time-dilation effect for the time interval of two events occurring at the same place. It is the time dilation effect that leads to $t_2 - t_1 = \gamma(t'_2 - t'_1) = \gamma T'$. Thus we have

$$T = t_{2r} - t_{1r} = (t_2 - t_1) + (R_2 - R_1)/c. \quad (1)$$

Using sine theorem in Fig. 3, we obtain

$$\frac{R_1}{\sin(\pi - \phi_2)} = \frac{R_2}{\sin \phi_1} = \frac{O'_1O'_2}{\sin(\phi_2 - \phi_1)}. \quad (2)$$

Taking advantage of Eq. (2) with $O'_1O'_2 = (t_2 - t_1)|\boldsymbol{\beta}|c$ taken into account, from Eq. (1) we have

$$T = (t_2 - t_1)\left[1 - |\boldsymbol{\beta}|\frac{\sin\phi_1 - \sin\phi_2}{\sin(\phi_1 - \phi_2)}\right]. \quad (3)$$

Inserting $t_2 - t_1 = \gamma T'$ into above with $T = 2\pi/\omega$ and $T' = 2\pi/\omega'$ employed, we obtain the Doppler formula for a spherical wave generated by a point light source, given by

$$\omega' = \omega\gamma\left[1 - |\boldsymbol{\beta}|\frac{\sin\phi_1 - \sin\phi_2}{\sin(\phi_1 - \phi_2)}\right], \quad (4)$$

where $\phi_1$ ($\phi_2$) is the position angle between the unit vector $\mathbf{n}_1$ ($\mathbf{n}_2$) and the velocity $\mathbf{v} = \boldsymbol{\beta}c$ measured by the observer at $t_{1r}$ ($t_{2r} = t_{1r} + T$).

Due to the relativity of motion, we can take the light source to be at rest while the observer moves at a velocity of $\mathbf{v}' = -\mathbf{v}$, as shown in Fig. 4. Considering that $T' = t'_2 - t'_1$, $t'_1 = t'_{1r} - R'_1/c$, $t'_2 = t'_{2r} - R'_2/c$, and $t'_{2r} - t'_{1r} = (t_2 - t_1)\gamma' = T\gamma'$ (time dilation), from a similar derivation we have

$$\omega = \omega'\gamma'\left[1 - |\boldsymbol{\beta}'|\frac{\sin\phi'_1 - \sin\phi'_2}{\sin(\phi'_1 - \phi'_2)}\right], \quad (5)$$

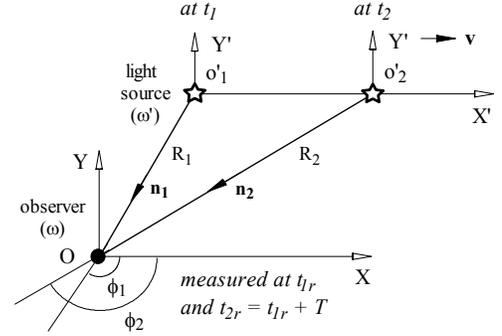

Fig. 3. A light source fixed in $X'O'Y'$ frame moves relatively to the observer fixed in $XOY$ frame at a velocity of $\mathbf{v} = \boldsymbol{\beta}c$ in the $x$-direction. Observed in the $XOY$ frame, the light source generates two consecutive crest-wavefronts at $t_1$ and $t_2$ respectively, and the observer receives them at the retarded times $t_{1r}$ and $t_{2r}$.

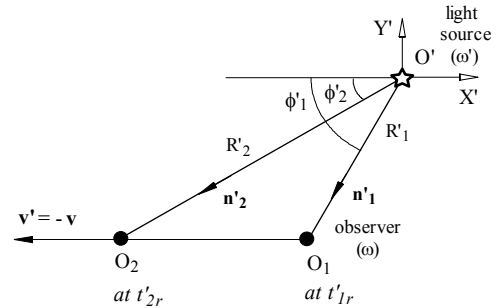

Fig. 4. The point light source fixed in $X'O'Y'$ frame is at rest, while the observer moves at a velocity of $\mathbf{v}' = -\mathbf{v}$ in the minus $x$-direction. Observed in the $X'O'Y'$ frame, the light source generates two consecutive crest-wavefronts at $t'_1$ and $t'_2$ respectively, and the moving observer receives them at the retarded times $t'_{1r}$ and $t'_{2r}$.

where $\phi_1'$ and $\phi_2'$ are the position angles between the unit wave vector $\mathbf{n}'$ and the velocity $\mathbf{v}' = \boldsymbol{\beta}'c = -\boldsymbol{\beta}c$, measured by an observer fixed with the light source at $t_1'$ and $t_2' = t_1' + T'$ respectively. Obviously, Eq. (4) and Eq. (5) reflects the principle of relativity.

Now let's take a look of the relations between the point-source and plane-wave Doppler effects. (1) When setting $\phi_1 = \phi_2 = 0$ or $\pi$ in Eq. (4), we have $\omega' = \omega\gamma(1 \mp \beta)$ with $\beta = |\boldsymbol{\beta}|$, which means that the point source and the plane wave have the same *conventional* longitudinal Doppler effect. (2) Letting $\phi_2 \to \phi_1$ or $\mathbf{n}_2 \to \mathbf{n}_1$, that is, the point source is set at infinity with respect to the observer, as supposed by Einstein [1], we obtain the Doppler formula for a plane wave, given by $\omega' = \omega\gamma(1 - \boldsymbol{\beta}\cdot\mathbf{n}_1)$. Therefore, application of the plane-wave Doppler formula to analysis of a moving point light source is a good approximation when the observer is far away from the light source [4].

To better understand the properties of the point-source Doppler effect, let's make some approximation analysis. It is seen from Fig. 3 that, $O_1'O_2' = \gamma T'\beta c = \gamma\beta\lambda'$ holds, with $\lambda' = cT'$ the proper wavelength of the moving light source. For $\gamma\beta\lambda' \ll R_1$, Eq. (4) can be approximated as

$$\frac{\lambda}{\lambda'} \approx \gamma(1 - \beta\cos\phi) + D_p, \quad \text{with} \quad \gamma\beta\lambda' \ll R \qquad (6)$$

where $R_1$ and $\phi_1$ are, respectively, replaced by $R$ and $\phi$, and

$$D_p = \frac{1}{2}\frac{\lambda'}{R}(\gamma\beta)^2\sin^2\phi. \qquad (7)$$

Note that the first term in Eq. (6) plays a role like a plane wave and the second term $D_p > 0$ is a red-shift modification caused by the point source, with $D_p$ depending on the proper wavelength $\lambda'$. $D_p = 0$ holds when $\phi = 0$ or $\pi$, while $D_p$ reaches maximum when $\phi = \pi/2$, suggesting that the transverse effect gets a maximum modification although the longitudinal effect is not affected, as mentioned above.

Physically, it is much easier to understand the relativistic effect when the Doppler formula is written in an approximate series of $\beta \ll 1$ [13,14]. Setting $\Delta\lambda \equiv \lambda - \lambda'$, from Eq. (6) we obtain a further simplified expression for the point-source Doppler formula

$$\Delta\lambda \approx \lambda'\left[(-\cos\phi)\beta + \left(\frac{1}{2} + \frac{1}{2}\frac{\lambda'}{R}\sin^2\phi\right)\beta^2\right]. \qquad (8)$$

In the above, the $\beta$-coefficient $(-\cos\phi)$ is the contribution of classical Doppler effect, while the $\beta^2$-coefficient has two parts: ½ comes from the relativistic effect, the same as for a plane wave, and $\lambda'\sin^2\phi/(2R)$ comes from a modification of the point source, both producing a red shift effect.

One of the ways to experimentally examine the relativistic effect is to determine the $\beta^2$-coefficient from a measured $\Delta\lambda$-*vs*-$\beta$ curve at a fixed $\phi$ for moving radiating atoms with a known transition frequency [14-17].

From Eq. (8) we can see that, to observe the point-source red-shift effect, it is necessary to directly measure the frequency of moving radiating atoms (ions) in the transverse direction. Such effect cannot be measured in the experiments by longitudinal observations [15-21], and those without directly measuring the frequency of the light re-emitted by the moving atoms (ions) [22-26].

Probably, the point-source red-shift effect may qualitatively explain why the $\beta^2$-coefficient is apparently larger by transverse observation in the previously-published research works: $0.498 \pm 0.025$ [16] and $0.491 \pm 0.017$ [17] both by longitudinal observation, while $0.52 \pm 0.03$ [14] by transverse observation (right angle), which is probably the only one so far, to our best knowledge.

It should be pointed out that, there is a "short-range" longitudinal Doppler effect for a moving point light source when the source is enough close to the observer ($\gamma\beta\lambda' \geq R_1$) so that $\phi_1 = 0$ and $\phi_2 = \pi$ are valid in Eq. (4) (see Appendix B).

## IV. CONCLUSIONS AND REMARKS

By means of a direct approach, we have derived the Doppler formula for a moving point light source, from which some conclusions result. (1) The point-source Doppler formula cannot be obtained from a standard Lorentz transformation, leading to an intrinsic Lorentz violation. (2) This formula contains an additional red-shift effect and a "short-range" longitudinal effect. (3) This formula is reduced into the one for a plane wave when the observer is far away from the source.

Traditionally, it has been generally understood for the principle of relativity that the mathematical equations expressing the laws of nature must be invariant in form under the Lorentz transformation (Lorentz invariance), and they must be Lorentz scalars, four-vectors, or four-tensors [3,6]; in other words, the principle of relativity and the Lorentz invariance are equivalent. However this is not true for the "wave four-vector $(\mathbf{k}, \omega/c)$" of the moving point light source (see Appendix A). From this we may conclude that the principle of relativity allows the existence of intrinsic Lorentz violation.

Theoretically the Doppler formula for a moving point light source may have some potential significance. (1) It clearly exposes in a primary, easy-to-understand level that the principle of relativity and the Lorentz invariance are not equivalent. (2) It indicates at what scale the intrinsic breaking of Lorentz invariance could be observed, helping in providing a guide for experimental test. Such a test could lead to an unexpected result that the frequency of a photon may not always keep constant in propagation (see Appendix C).

Finally, we would like to make some remarks on Doppler effect. From a moving frame to the lab frame, EM fields can be obtained from Lorentz transformation of field-strength tensors [3]; however, the transformation of frequency or Doppler frequency shift needs additional calculations based on invariance of phase and the principle of relativity, and the derivation of Doppler formula only needs the phase function, without a need of knowing the EM field amplitudes.

In the point-source Doppler derivation, the wave period, observed in the lab frame, is taken as a primary quantity, while the frequency is a derived quantity. That is because if the frequency were taken to be the primary quantity (instead of the period), it would be difficult to set up the steps about how to measure the frequency. Obviously, this process is different from that given in traditional textbooks [3], where the frequency is taken as a primary quantity, because it is usually supposed to be known or not to change with time and position.

The wave-period definition used in the paper is a generalization from the previous analysis of plane-wave Doppler effect [4]. When the observer is far away from the point source, this Doppler formula is reduced back to the one for a plane wave, which is consistent with commonly-used correspondence principle.

One might question: Is the point-source Doppler formula, Eq. (4), compatible with Maxwell equations? The answer is "yes", which is shown as follow.

Suppose the point-source field solution in the source-rest frame is given by $\mathbf{A}'\exp i(\omega't'-|\mathbf{k}'||\mathbf{x}'|) = \mathbf{A}'\exp i\omega'(t'-|\mathbf{x}'|/c)$. After time-space Lorentz transformations $t'=t'(\mathbf{x},t)$ and $\mathbf{x}'=\mathbf{x}'(\mathbf{x},t)$, and EM field-strength tensor transformation with the amplitude $\mathbf{A}' \to \mathbf{A}$, we have the solution in the lab frame, given by $\mathbf{A}\exp i\omega'[t'(\mathbf{x},t)-|\mathbf{x}'(\mathbf{x},t)|/c)]$, which also satisfies Maxwell equations in the lab frame. The frequency $\omega$, observed in the lab frame, is governed by Eq. (4) under the condition of $\omega'=\omega'(\omega,\mathbf{x},t)=$ constant; thus there is nothing changed mathematically for the solution to satisfy Maxwell equations, although the forms of the two phase functions become different. Therefore, the point-source Doppler formula satisfies all the conditions required by the relativistic electrodynamics: (1) principle of relativity, (2) constancy of light speed, (3) invariance of phase, and (4) being compatible with Maxwell equations.

## ACKNOWLEDGMENT

The author would like to thank Dr. Andrew M. Sessler of LBNL for his helpful discussions and critical comments.

## APPENDIX A: WHY IS THE LORENTZ COVARIANCE OF $(\mathbf{k}',\omega'/c)$ VIOLATED FOR A MOVING POINT LIGHT SOURCE?

In this Appendix we will show that, if the wave vector and frequency for a moving point source were to follow Lorentz transformation, an unphysical phenomenon could result; in other words, the covariance of $(\mathbf{k}',\omega'/c)$ cannot hold in such a case.

For a spherical wave in free space, generated from a moving point source that is fixed at the origin ($\mathbf{x}'=0$) of $X'Y'Z'$ frame, the phase function is given by $\Psi'=\omega't'-|\mathbf{k}'||\mathbf{x}'|$, with $|\mathbf{k}'|=\omega'/c$. To reflect the constraint between $(\mathbf{x}',ct')$ and $(\mathbf{k}',\omega'/c)$ in an inner-product form, $\Psi'$ can be written as

$$\Psi'=\omega't'-\mathbf{k}'_p\cdot\mathbf{x}', \qquad \text{with}\qquad \mathbf{k}'_p=\frac{\omega'}{c}\frac{\mathbf{x}'}{|\mathbf{x}'|}. \qquad (A1)$$

If we define $(\mathbf{k}'_p,\omega'/c)$ as a Lorentz covariant four-vector, then the invariance of phase is automatically satisfied, namely $\Psi'=\omega't'-|\mathbf{k}'||\mathbf{x}'|=g'_{\mu\nu}K'^{\mu}X'^{\nu}$, with $K'^{\mu}=(\mathbf{k}'_p,\omega'/c)$ and the Minkowski metric $g'_{\mu\nu}=diag(-1,-1,-1,+1)$. However, it should be noted that because $(\mathbf{k}'_p,\omega'/c)$ and $(\mathbf{x}',ct')$ are not independent, the covariance of $(\mathbf{k}'_p,\omega'/c)$ is only a *sufficient* condition for the invariance of phase, instead of a *sufficient and necessary* condition. Thus there are two options about how to treat $(\mathbf{k}'_p,\omega'/c)$: (*a*) set $(\mathbf{k}'_p,\omega'/c)$ to follow Lorentz transformation, and (*b*) set $(\mathbf{k}'_p,\omega'/c)$ not to follow Lorentz transformation. The two options are both allowable mathematically. In Sec. III, option (*b*) is taken. The two options produce the same phase function and the same *unit* wave vector in the lab frame, but different Doppler formulas. In the following, we will show that option (*a*) contains an unphysical Doppler frequency shift and it should be discarded.

Suppose that the source-rest frame $X'Y'Z'$ moves at $\boldsymbol{\beta}$ with respect to the lab frame $XYZ$. Following option (*a*), we have $\Psi'=g'_{\mu\nu}K'^{\mu}X'^{\nu}$. The Lorentz transformations of $X^{\mu}=(\mathbf{x},ct)$ is given by [3]

$$\mathbf{x}'=\mathbf{x}+\frac{\gamma-1}{\beta^2}(\boldsymbol{\beta}\cdot\mathbf{x})\boldsymbol{\beta}-\gamma\boldsymbol{\beta}ct, \qquad (A2)$$

$$ct'=\gamma(ct-\boldsymbol{\beta}\cdot\mathbf{x}). \qquad (A3)$$

With Eq. (A1) taken into account, the Lorentz transformation of $K^{\mu}=(\mathbf{k}_p,\omega/c)$ from $K'^{\mu}=(\mathbf{k}'_p,\omega'/c)$ is given by

$$\mathbf{k}_p=\frac{\omega'}{c}\left(\frac{\mathbf{x}'}{|\mathbf{x}'|}+\frac{\gamma-1}{\beta^2}\boldsymbol{\beta}\cdot\frac{\mathbf{x}'}{|\mathbf{x}'|}\boldsymbol{\beta}+\gamma\boldsymbol{\beta}\right), \qquad (A4)$$

$$\frac{\omega}{c}=\frac{\omega'}{c}\gamma\left(1+\boldsymbol{\beta}\cdot\frac{\mathbf{x}'}{|\mathbf{x}'|}\right), \qquad \text{(Doppler formula)} \qquad (A5)$$

where $\mathbf{x}'$ is given by Eq. (A2). Note: the unit wave vector is given by $\hat{\mathbf{n}}'=\mathbf{k}'_p/|\mathbf{k}'_p|=\mathbf{x}'/|\mathbf{x}'|$ in the source-rest frame, while it is given by $\hat{\mathbf{n}}=\mathbf{k}_p/|\mathbf{k}_p|$ in the lab frame.

It is seen from Eq. (A5) that, observed in the lab frame, the frequency $\omega$ is not determinate when the observer and the point source overlap ($\mathbf{x}'=0$), which is *not physical*. Thus the covariance of $(\mathbf{k}'_p,\omega'/c)$ cannot hold. However, it can be shown that $\mathbf{n}_1=\mathbf{R}_1/R_1$ in Fig. (3) is equal to $\hat{\mathbf{n}}=\mathbf{k}_p/|\mathbf{k}_p|$, namely, options (*a*) and (*b*) have the same unit wave vector in the lab frame, as mentioned before.

## APPENDIX B: SHORT-RANGE LONGITUDINAL DOPPLER EFFECT

In this Appendix, we will show that, there is a "short-range" longitudinal Doppler effect for a spherical wave when the point light source is so close to the observer that $\phi_1=0$ and $\phi_2=\pi$ hold in Eq. (4).

As sown in Fig. B1, the point light source emits the first and second crest-wavefronts at $(t_1,O'_1)$ and $(t_2,O'_2)$ respectively, with $O'_1O'_2=(t_2-t_1)\beta c=\gamma T'\beta c=\gamma\beta\lambda'$. When $O'_1$ and $O'_2$ both fall between $A$ and $B$, with $AO=OB=O'_1O'_2$ ($\phi_1=0$ and $\phi_2=\pi$), we have

$$\xi=\frac{\sin\phi_1-\sin\phi_2}{\sin(\phi_1-\phi_2)}=2\frac{O'_1O}{AO}-1. \qquad (B1)$$

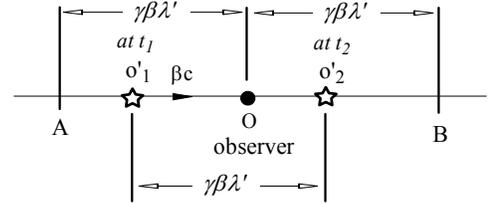

Fig. B1. Illustration of short-range longitudinal Doppler effect. When $O'_1$ and $O'_2$ both fall between $A$ and $B$, we have $1>\xi>-1$ holding; otherwise, $\xi=1$ for both $O'_1$ and $O'_2$ on the left of $O$, and $\xi=-1$ for both $O'_1$ and $O'_2$ on the right of $O$.

Accordingly, we have three cases for the longitudinal Doppler effect in Eq. (4). (*i*) Up-shift effect: $\omega=\omega'[(1+\beta)/(1-\beta)]^{1/2}$ for $\xi=1$, with both $O'_1$ and $O'_2$ on the left of $O$ ($\phi_1=\phi_2=0$). (*ii*) Short-range effect: $\omega=\omega'/[\gamma(1-\beta\xi)]$ for $1>\xi>-1$, with both $O'_1$ and $O'_2$ between $A$ and $B$ ($\phi_1=0$ and $\phi_2=\pi$). (*iii*) Down-shift effect: $\omega=\omega'[(1-\beta)/(1+\beta)]^{1/2}$ for $\xi=-1$, with both $O'_1$ and $O'_2$ on the right of $O$ ($\phi_1=\phi_2=\pi$).

The zero-shift condition in such a case can be obtained by solving $\gamma(1-\beta\xi)=1$. With $\xi=(\gamma-1)^{1/2}(\gamma+1)^{-1/2}$ inserted into Eq. (B1) we have

$$\frac{O'_1O}{AO}=\frac{1}{2}\left(1+\sqrt{\frac{\gamma-1}{\gamma+1}}\right). \qquad (B2)$$

In other words, the time interval of the observer's receiving two consecutive crest-wavefronts emitted at $O_1'$ and $O_2'$, which satisfy the above Eq. (B2), is equal to the proper time interval, namely $t_{2r} - t_{1r} = t_2' - t_1'$ or $T = T'$.

For the short-range Doppler effect produced when the point source moves from $A$ to $B$, the measured frequency versus the source frequency varies continuously in the range of

$$\sqrt{\frac{1+\beta}{1-\beta}} > \frac{\omega}{\omega'} > \sqrt{\frac{1-\beta}{1+\beta}} \ . \qquad (B3)$$

As it is well known from university physics textbooks [11], for a moving point light source there is a jump between the longitudinal Doppler up- and down-shifts calculated from the plane wave formula [1], while they are continuous from Eq. (4). That is because the plane wave formula is only applicable to the case where the observer is far away from the source. For example, when the observer overlaps with the point source, the plane wave formula cannot give a determinate value due to the indetermination of the position angle $\phi$ [4], while Eq. (4) gives a unique value, $\omega = \omega'[(1-\beta)/(1+\beta)]^{1/2}$, with $\phi_2 = \pi$, leading to $\xi = -1$, no matter what $\phi_1$ is.

## APPENDIX C: SUGGESTED SCHEME OF EXPERIMENT FOR POINT-SOURCE DOPPLER EFFECT

Laser saturation spectroscopy has been successfully used to confirm Einstein's Doppler formula with unprecedented precision, as reported in previously-published research works [21,24,25]. In the experiments by the authors, the frequencies of two anti-parallel propagating lasers are adjusted to reach Doppler-resonance with the transition frequency of moving ions. But the frequency of the light emitted by the ions is not measured in the transverse direction, as stated in the Comment [26], although they put a recording of the number of photons to monitor Lamb dip. Based on their experiments, a conceptual scheme to experimentally test the Doppler formula for a moving point light source is proposed here, as shown in Fig. C1.

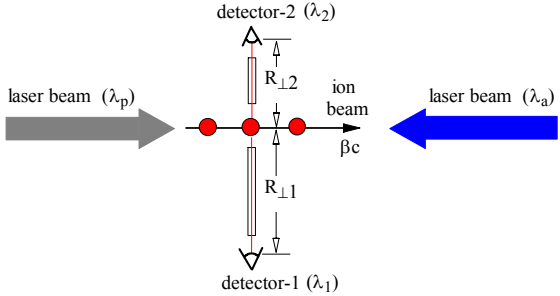

Fig. C1. Conceptual experimental scheme to test Doppler formula for a moving point light source by laser saturation spectroscopy. Two anti-parallel propagating lasers with wavelengths $\lambda_p$ and $\lambda_a$ are adjusted to reach resonance with a moving ion beam so that the transition wavelength $\lambda_0 = (\lambda_p \lambda_a)^{1/2}$. The frequency of fluorescent light emitted by the ions is observed in two symmetric transverse directions with different distances, $R_{\perp 1}$ and $R_{\perp 2}$ respectively, and with measured wavelengths $\lambda_1$ and $\lambda_2$.

It is seen from Fig. C1 that, the frequency of fluorescent light emitted by the moving ions, which correspond to identical point light sources, are measured in two symmetric transverse directions, with one transverse distance lager than the other.

From Eq. (8), the Doppler shift formula in such a case is given by

$$\Delta\lambda \approx \lambda_0 \left(\frac{1}{2} + \frac{1}{2}\frac{\lambda_0}{R_\perp}\right)\beta^2, \quad \text{with } \beta\lambda_0 << R_\perp \qquad (C1)$$

where $\beta << 1$, $\Delta\lambda \equiv \lambda - \lambda_0$ with $\lambda_0$ the ion transition wavelength (namely point-source proper wavelength) and $\lambda$ the measured wavelength in the transverse direction, and $R_\perp$ is the transverse distance, as shown in Fig. C1. The term $\lambda_0/(2R_\perp)$ is resulting from the point-source red-shift modification, as indicated in Sec. III, and the shift $\Delta\lambda$ is reduced as the increase of $R_\perp$. If $\lambda_1 < \lambda_2$ is observed for $R_{\perp 1} > R_{\perp 2}$, then the point-source red-shift effect, or the intrinsic Lorentz violation will be confirmed, qualitatively at least.

It is worthwhile to point out that in the laser saturation spectroscopy, no matter whether one transition [21,25] or two transitions [24] are driven, the Doppler effect is confirmed for the moving ion as an observer who takes the light from lasers to be "local plane waves", because the ion's dimension is much smaller than the laser-beam size; the very ion-observer tells the experimenter what the lasers' frequency is, that he observed. To verify the point-source Doppler effect, a direct measurement of the light emitted by the moving ion is required, namely the experimenter must be "a real observer".

A striking prediction of Eq. (C1) is that the observed frequency of photons emitted by moving ions changes with the transverse distance $R_\perp$; thus challenging the constancy of Planck constant in Einstein's light-quantum hypothesis $E = \hbar\omega$ since the energy conservation must hold. From Eq. (C1), we may obtain its relative change, given by

$$\frac{\hbar}{\hbar_\infty} \approx 1 + \frac{1}{2}\frac{\lambda_0}{R_\perp}\beta^2, \text{ with } \beta << 1 \text{ and } \beta\lambda_0 << R_\perp \qquad (C2)$$

where $\hbar = \hbar(R_\perp)$ depends on $R_\perp$, and $\hbar_\infty \equiv \hbar(R_\perp = \infty)$ equal to the plane-wave Planck constant [27], because the spherical wave behaves as a plane wave observed at infinity. Such a prediction sounds unacceptable, but it is a strict result of the principle of relativity, just like the red shift for approaching [4].

As a theoretical interest, one might ask: What is the Planck constant when a photon is just leaving the point source? This can be evaluated from the following analysis.

From Eq. (4), we know that the observed frequency is $\omega = \omega'[(1-\beta)/(1+\beta)]^{1/2}$ when the observer overlaps the source (with $R_\perp = 0$ and $\phi_2 = \pi$), while the frequency is $\omega = \omega'/\gamma(1-\beta\cos\phi)$ observed at infinity ($\phi_2 \to \phi_1 = \phi$, the same as that for a plane wave). The photon's momentum $\mathbf{p} = (\hbar\omega/c)\hat{\mathbf{n}}$ and the energy $E = \hbar\omega$ are conserved during the propagation and they constitute a Lorentz four-vector $P^\mu = (\mathbf{p}, E/c)$, where $\hat{\mathbf{n}} \equiv \mathbf{n}_1 = \mathbf{R}_1/R_1$ is the unit wave vector [confer Fig. (3)]. From the energy conservation, we obtain

$$\frac{\hbar_\phi}{\hbar_\infty} = \frac{1+\beta}{1-\beta\cos\phi}, \qquad (C3)$$

where $\hbar_\phi$ is the Planck "constant" for the photon with an emitting angle of $\phi$ with respect to the source moving direction, and $\hbar_\infty \equiv \hbar(R = \infty)$ is the plane-wave Planck constant as mentioned above. $\hbar_\phi$ is a real constant when the source is at rest ($\beta = 0$), but it is not when the source moves, because the just-leaving-source photons with different emission angles have different energies while they have the same frequency.

According to the above analysis, the Planck constant for a moving point source is a real constant observed in the source-rest frame, but it is not a constant observed in the lab frame. Thus the Planck constant for a moving point source is not a Lorentz invariant constant or universal constant; nevertheless, it is an approximate universal constant when the observer is far away from the source or the source moves slowly enough.

In contrast, as shown in [27], the Planck constant for plane waves is exactly a universal constant.

---
The following three attachments give more details for Appendix A. A 3-space-notation proof of the violation of Lorentz covariance for a moving point source is given in Attachment-I, while the 4-space-notation proof is given in Attachment-II. In Attachment-III, the unit wave vector, observed in the lab frame, is derived by use of Lorentz transformation.

## Attachment-I: Why is the Lorentz covariance of $(\mathbf{k}', \omega'/c)$ violated for a moving point light source? (In 3-space notation)

Some scientists in the community insist that Einstein's plane wave Doppler formula should be applicable to a moving point light source, no matter whether the observer is close to the source or not. A strong argument is that "the plane wave decomposition is mathematically universal", and the spherical wave produced by the moving point source can be decomposed into plane waves. At first thought, this argument sounds correct, but on second thoughts, it is questionable. Why?

--------

**A note.** A *physical* plane wave in free space is defined as such a plane wave that can exist independently. Einstein's Doppler formula is applicable to any physical plane waves in free space. Many time-harmonic EM fields can be decomposed into a sum of "plane waves", but the component plane waves are not necessarily *physical* plane waves; the ones for a point-source EM field, for example, which is shown below.

The spherical wave produced by a rest point light source located at $\mathbf{x} = \mathbf{x}'$, namely Green's function, satisfies wave equation

$$(\nabla^2 + \mathbf{k}_0^2) G(\mathbf{x}, \mathbf{x}') = -\delta(\mathbf{x} - \mathbf{x}'), \qquad \text{with} \qquad |\mathbf{k}_0| = \frac{\omega_0}{c}.$$

where $\omega_0$ is the frequency of the given point light source. By taking advantage of the 3D delta-function expression

$$\delta(\mathbf{x} - \mathbf{x}') = \left(\frac{1}{2\pi}\right)^3 \iiint e^{-i\mathbf{k}\cdot(\mathbf{x}-\mathbf{x}')} d^3\mathbf{k},$$

the plane-wave decomposition of the spherical wave can be written as a singular Fourier integral [J. D. Jackson, Classical Electrodynamics, 3rd edition, (John Wiley & Sons, NJ, 1999), Chapter 12, p. 612, Eq. (12.129); also D. J. Griffiths, Introduction to quantum mechanics, (Prentice Hall, NJ, 1995), p. 364, Eq. (11.48)]

$$G(\mathbf{x}, \mathbf{x}') = \frac{e^{-i|\mathbf{k}_0||\mathbf{x}-\mathbf{x}'|}}{4\pi|\mathbf{x}-\mathbf{x}'|}(\text{spherical wave}) = \left(\frac{1}{2\pi}\right)^3 \iiint \frac{e^{-i\mathbf{k}\cdot(\mathbf{x}-\mathbf{x}')}(\text{plane wave factor})}{\mathbf{k}^2 - \mathbf{k}_0^2} d^3\mathbf{k},$$

which converges in the sense of taking a limit for a designated contour in the *k*- complex plane. Thus the component plane wave has a *complex wave vector* $\mathbf{k}$, and this is not a physical plane wave, because it *cannot exist independently*. A plane wave with a complex wave number in free space is not consistent with energy conservation law in the sense of classical electrodynamics, and it is also not consistent with photon momentum hypothesis ($\mathbf{p} = \hbar \mathbf{k}$) in the sense of quantum mechanics. Therefore, **the plane-wave decomposition of a point-source-generated spherical wave is not a physical-plane-wave decomposition**, just a kind of mathematical correspondence (treatment).

In contrast, the spherical-wave decomposition of a plane wave is a sum of *physical* spherical waves, given by [J. D. Jackson, Classical Electrodynamics, 3rd edition, (John Wiley & Sons, NJ, 1999), p. 471, Eq. (10.44) in Chapter 10; also D. J. Griffiths, Introduction to quantum mechanics, (Prentice Hall, NJ, 1995), p. 361, Eq. (11.30)]

$$e^{i\mathbf{k}\cdot\mathbf{x}} = \sum_{l=0}^{+\infty} i^l (2l+1) j_l(|\mathbf{k}||\mathbf{x}|) P_l(\cos\theta),$$

where $\theta$ is the angle between $\mathbf{k}$ and $\mathbf{x}$. The spherical-wave factor $j_l(|\mathbf{k}||\mathbf{x}|)P_l(\cos\theta)$, analytical at $|\mathbf{x}| = 0$, is the azimuthally symmetric spherical harmonic function of the first kind, and it is denotes a *physical* standing wave in the radial direction. The lowest-order term $j_0(|\mathbf{k}||\mathbf{x}|)P_0(\cos\theta)$ has a completely spherical symmetry, since the zero[th]-order Legendre function $P_0(\cos\theta) = 1$. Each of the component spherical waves in the sum is physical and it *can exist independently*.

A finite *physical* plane-wave series decomposition of spherical waves is given by MacPhie and Ke-Li Wu, "A Plane Wave Expansion of Spherical Wave Functions for Modal Analysis of Guided Wave Structures and Scatterers", IEEE Trans. Antennas and Propagation **51**, 2801 (2003). Note: These spherical waves are analytical at $|\mathbf{x}| = 0$, name they are spherical harmonics functions of the first kind.

--------

(Let us put aside whether the spherical wave can be decomposed into physical plane waves.) Suppose that the spherical wave can be decomposed into plane waves; however, we still cannot obtain the Doppler frequency shift of the whole spherical wave from individual component plane waves, because all component plane waves propagate in *all different directions* (otherwise not a spherical wave), while Doppler effects depend on individual plane-wave propagation directions. Thus how to define the wave period or frequency of the whole spherical wave becomes questionable. Therefore, the plane-wave decomposition has no help in solving the problem.

From a moving frame to the lab frame, EM fields can be obtained from Lorentz transformation of field-strength tensors; however, the transformation of frequency or Doppler frequency shift needs additional derivations based on invariance of phase and the principle of relativity, and the derivation of Doppler formula only needs the phase function.

**Plane wave**: According to the principle of relativity, the phase factor $\exp i(\omega t - \mathbf{k} \cdot \mathbf{x})$ of a plane wave in free space is symmetric with respect to all inertial frames. $(\mathbf{x}, ct)$ and $(\mathbf{k}, \omega/c)$ are *completely independent*. Since $X^\mu = (\mathbf{x}, ct)$ must be a Lorentz covariant 4-vector, the invariance of phase $\Psi = (\omega t - \mathbf{k} \cdot \mathbf{x})$ and the covariance of $(\mathbf{k}, \omega/c)$ are equivalent, as shown by Einstein in 1905; that is, the invariance of phase must result in the covariance of $(\mathbf{k}, \omega/c)$. Thus writing $K^\mu = (\mathbf{k}, \omega/c)$, we have $\Psi = g_{\mu\nu} K^\mu X^\nu = (\omega t - \mathbf{k} \cdot \mathbf{x})$, with the metric tensor $g_{\mu\nu} = g^{\mu\nu} = diag(-1,-1,-1,+1)$.

**Conclusion:** For a plane wave with $\Psi = (\omega t - \mathbf{k} \cdot \mathbf{x})$,
 (1) $(\mathbf{x}, ct)$ and $(\mathbf{k}, \omega/c)$ are completely independent;
 (2) $K^\mu = (\mathbf{k}, \omega/c)$ must follow Lorentz transformation, with *no other option*.

**Moving point source**: For a spherical wave in free space, generated from a moving point light source that is fixed at the origin ($\mathbf{x}' = 0$) of $X'Y'Z'$ frame, the phase function is given by $\Psi' = \omega' t' - |\mathbf{k}'||\mathbf{x}'|$ in the source-rest frame $X'Y'Z'$, with $|\mathbf{k}'| = \omega'/c$, and it is also a Lorentz invariant. But there is an additional strong constraint between $(\mathbf{x}', ct')$ and $(\mathbf{k}', \omega'/c)$; $\mathbf{k}' \cdot \mathbf{x}' = |\mathbf{k}'||\mathbf{x}'|$ must hold. To reflect the constraint in an inner-product manner, the phase function $\Psi' = \omega' t' - |\mathbf{k}'||\mathbf{x}'|$ can be written as

$$\Psi' = \omega' t' - |\mathbf{k}'||\mathbf{x}'| = \omega' t' - \left( |\mathbf{k}'| \frac{\mathbf{x}'}{|\mathbf{x}'|} \right) \cdot \mathbf{x}' = \omega' t' - \mathbf{k}'_p \cdot \mathbf{x}' \tag{I-1}$$

where, to reflect the constraint between $(\mathbf{x}', ct')$ and $(\mathbf{k}', \omega'/c)$, the point-source wave vector is written as

$$\mathbf{k}'_p = |\mathbf{k}'| \frac{\mathbf{x}'}{|\mathbf{x}'|} = \frac{\omega'}{c} \frac{\mathbf{x}'}{|\mathbf{x}'|} . \tag{I-2}$$

If we define $K'^\mu = (\mathbf{k}'_p, \omega'/c)$ as a Lorentz covariant 4-vector, then the invariance of the phase $\Psi' = \omega' t' - |\mathbf{k}'||\mathbf{x}'| = g'_{\mu\nu} K'^\mu X'^\nu$ is automatically satisfied. However, it should be emphasized that because $(\mathbf{k}'_p, \omega'/c)$ is not independent of $(\mathbf{x}', ct')$, the covariance of $(\mathbf{k}'_p, \omega'/c)$ is only a sufficient condition for the invariance of phase, instead of a sufficient and necessary condition. Just because of this, there are two options about how to treat $(\mathbf{k}'_p, \omega'/c)$: (*a*) Make $(\mathbf{k}'_p, \omega'/c)$ follow Lorentz transformation, and (*b*) Make $(\mathbf{k}'_p, \omega'/c)$ not follow Lorentz transformation. The two options result in the same phase function $\Psi(\omega'; \mathbf{x}, t)$, and the unit wave vector in the lab frame (see Attachment-III), but produce different Doppler formulas. [Note: Since $\omega'$ is the point-source proper frequency and it is not frame-dependent, it is easy to show mathematically that $\omega' t' - |\mathbf{k}'||\mathbf{x}'| = \omega'(t' - |\mathbf{x}'|/c) = \Psi' = $ constant is always a spherical surface under time-space Lorenz transformation; observed in the source-rest frame $X'Y'Z'$, the center of the spherical surface is at the $X'Y'Z'$'s origin $\mathbf{x}' = 0$, while observed in the $XYZ$ frame which moves with respect to $X'Y'Z'$ frame, the center is away from the $XYZ$'s origin $\mathbf{x} = 0$. Because of the invariance of phase, no matter what options for $(\mathbf{k}'_p, \omega'/c)$ are taken, the resultant phase function is the same, just producing different Doppler formulas.]

**Conclusion:** For a moving point light source with $\Psi' = \omega' t' - |\mathbf{k}'||\mathbf{x}'| = \omega' t' - \mathbf{k}'_p \cdot \mathbf{x}'$,
 (1) $(\mathbf{x}', ct')$ and $(\mathbf{k}'_p, \omega'/c)$ are not independent;
 (2) There are *two options* to treat $(\mathbf{k}'_p, \omega'/c)$: (*a*) following Lorentz transformation, (*b*) not following Lorentz transformation.

In this paper, option (*b*) is taken. In the following, we will show that option (*a*) will results in an unphysical result and it should be discarded.

Suppose that the source-rest frame $X'Y'Z'$ moves at $\boldsymbol{\beta}$ with respect to the lab frame $XYZ$. Following option (*a*), we have $\Psi' = g'_{\mu\nu} K'^\mu X'^\nu$, with $K'^\mu = (\mathbf{k}'_p, \omega'/c)$ and $X'^\mu = (\mathbf{x}', ct')$. The Lorentz transformations of $X^\mu = (\mathbf{x}, ct)$ is given by [J. D. Jackson, Classical Electrodynamics, 3$^{rd}$ edition, (John Wiley & Sons, NJ, 1999), p. 525, Eq. (11.19)]

$$\mathbf{x} = \mathbf{x}' + \frac{\gamma - 1}{\beta'^2} (\boldsymbol{\beta}' \cdot \mathbf{x}') \boldsymbol{\beta}' - \gamma \boldsymbol{\beta}' ct', \tag{I-3}$$

$$ct = \gamma(ct' - \boldsymbol{\beta}' \cdot \mathbf{x}') . \tag{I-4}$$

With $\boldsymbol{\beta}' = -\boldsymbol{\beta}$, the inverse transformation is given by

$$\mathbf{x}' = \mathbf{x} + \frac{\gamma - 1}{\beta^2}(\boldsymbol{\beta} \cdot \mathbf{x})\boldsymbol{\beta} - \gamma \boldsymbol{\beta} ct, \qquad (I\text{-}5)$$

$$ct' = \gamma(ct - \boldsymbol{\beta} \cdot \mathbf{x}). \qquad (I\text{-}6)$$

With Eq. (I-2) taken into account, the Lorentz transformations of $K^\mu = (\mathbf{k}_p, \omega/c)$ is given by

$$\mathbf{k}_p = \mathbf{k}'_p + \frac{\gamma-1}{\beta^2}(\boldsymbol{\beta}' \cdot \mathbf{k}'_p)\boldsymbol{\beta}' - \gamma \boldsymbol{\beta}'\left(\frac{\omega'}{c}\right) = \frac{\omega'}{c}\left(\frac{\mathbf{x}'}{|\mathbf{x}'|} + \frac{\gamma-1}{\beta^2}\boldsymbol{\beta}' \cdot \frac{\mathbf{x}'}{|\mathbf{x}'|}\boldsymbol{\beta}' - \gamma \boldsymbol{\beta}'\right), \qquad (I\text{-}7)$$

$$\frac{\omega}{c} = \gamma\left(\frac{\omega'}{c} - \boldsymbol{\beta}' \cdot \mathbf{k}'_p\right) = \frac{\omega'}{c}\gamma\left(1 - \boldsymbol{\beta}' \cdot \frac{\mathbf{x}'}{|\mathbf{x}'|}\right), \qquad \text{(Doppler formula)} \qquad (I\text{-}8)$$

and the phase function is given by

$$\Psi = g_{\mu\nu} K^\mu X^\nu = \omega t - \mathbf{k}_p \cdot \mathbf{x}. \qquad (I\text{-}9)$$

The above covariant form clearly shows $\Psi = \Psi'$. From Eq. (I-5), $\mathbf{x}'/|\mathbf{x}'|$ in Eqs. (I-7) and (I-8) is given by

$$\frac{\mathbf{x}'}{|\mathbf{x}'|} = \frac{\mathbf{x} + \dfrac{\gamma-1}{\beta^2}(\boldsymbol{\beta} \cdot \mathbf{x})\boldsymbol{\beta} - \gamma \boldsymbol{\beta} ct}{\left|\mathbf{x} + \dfrac{\gamma-1}{\beta^2}(\boldsymbol{\beta} \cdot \mathbf{x})\boldsymbol{\beta} - \gamma \boldsymbol{\beta} ct\right|}. \qquad (I\text{-}10)$$

Note: In the source-rest frame, $\mathbf{k}'_p \,/\!/\, \mathbf{x}'$ holds, while in the lab frame, $\mathbf{k}_p \,/\!/\, \mathbf{x}$ usually does not hold. In such a case, the unit wave vector in the lab frame can be obtained from Eq. (I-7)/Eq. (I-8), given by

$$\hat{\mathbf{n}} = \frac{\mathbf{k}_p}{\omega/c} = \frac{\dfrac{\mathbf{x}'}{|\mathbf{x}'|} + \dfrac{\gamma-1}{\beta^2}\boldsymbol{\beta}' \cdot \dfrac{\mathbf{x}'}{|\mathbf{x}'|}\boldsymbol{\beta}' - \gamma\boldsymbol{\beta}'}{\gamma\left(1 - \boldsymbol{\beta}' \cdot \dfrac{\mathbf{x}'}{|\mathbf{x}'|}\right)} = \frac{\dfrac{\mathbf{x}'}{|\mathbf{x}'|} + \dfrac{\gamma-1}{\beta^2}\boldsymbol{\beta} \cdot \dfrac{\mathbf{x}'}{|\mathbf{x}'|}\boldsymbol{\beta} + \gamma\boldsymbol{\beta}}{\gamma\left(1 + \boldsymbol{\beta} \cdot \dfrac{\mathbf{x}'}{|\mathbf{x}'|}\right)} \qquad (I\text{-}11)$$

So far we have obtained the Lorentz transformation of $K^\mu = (\mathbf{k}_p, \omega/c)$. It is seen from Eq. (I-8) that, (i) the observed frequency $\omega$ changes with time and location, which is against traditional concepts; (ii) like the Einstein's Doppler formula, the frequency is not determinate when the observer and the point source overlap ($\mathbf{x}' = 0$), which is NOT physical.

The math derivations, $(\mathbf{k}', \omega'/c) \to (\mathbf{k}'_p, \omega'/c)$ from Eqs. (I-1) and (I-2), and $(\mathbf{k}'_p, \omega'/c) \to (\mathbf{k}_p, \omega/c)$ from Eqs. (I-7) and (I-8), are all strict, and there is no any component of the derivations open to question.

If the frequency changes during photon's propagation, the Planck constant also should change to keep the energy conservation law valid.

It is widely assumed that the Planck constant is a Lorentz invariant (universal constant); interestingly, a math proof of the invariance for plane waves is given in http://arxiv.org/abs/1106.1163 .

**Conclusion:** Option (*a*) is not physical, namely the wave vector $\mathbf{k}'$ and the frequency $\omega'$ for a moving point light source cannot not constitute a Lorentz covariant 4-vector.

**Attachment-II: Why is the Lorentz covariance of $(\mathbf{k}', \omega'/c)$ violated for a moving point light source? (In 4-space notation)**

For the sake of simplification, we suppose that $X'Y'Z'$ frame moves with respect to $XYZ$ frame only in the $x$-direction, namely $\boldsymbol{\beta} = \hat{\mathbf{x}}\beta$. The time-space Lorentz transformation is given by

$$\begin{pmatrix} x' \\ y' \\ z' \\ ct' \end{pmatrix} = \begin{pmatrix} \gamma & 0 & 0 & -\gamma\beta \\ 0 & 1 & 0 & 0 \\ 0 & 0 & 1 & 0 \\ -\gamma\beta & 0 & 0 & \gamma \end{pmatrix} \begin{pmatrix} x \\ y \\ z \\ ct \end{pmatrix}, \text{ or } \begin{pmatrix} x \\ y \\ z \\ ct \end{pmatrix} = \begin{pmatrix} \gamma & 0 & 0 & -\gamma\beta' \\ 0 & 1 & 0 & 0 \\ 0 & 0 & 1 & 0 \\ -\gamma\beta' & 0 & 0 & \gamma \end{pmatrix} \begin{pmatrix} x' \\ y' \\ z' \\ ct' \end{pmatrix}, \tag{II-1}$$

corresponding to $X'^{\mu} = \Lambda^{\mu}_{\nu} X^{\nu}$ or $X^{\mu} = \Lambda'^{\mu}_{\nu} X'^{\nu}$, where $\beta' = -\beta$. The Lorentz transformation matrices are given by

$$\Lambda = \begin{pmatrix} \gamma & 0 & 0 & -\gamma\beta \\ 0 & 1 & 0 & 0 \\ 0 & 0 & 1 & 0 \\ -\gamma\beta & 0 & 0 & \gamma \end{pmatrix}, \qquad \Lambda' = \begin{pmatrix} \gamma & 0 & 0 & -\gamma\beta' \\ 0 & 1 & 0 & 0 \\ 0 & 0 & 1 & 0 \\ -\gamma\beta' & 0 & 0 & \gamma \end{pmatrix}, \tag{II-2}$$

with $\Lambda^T = \Lambda$ and $(\Lambda')^T = \Lambda'$, and they satisfy

$$\Lambda\Lambda' = \begin{pmatrix} \gamma & 0 & 0 & -\gamma\beta \\ 0 & 1 & 0 & 0 \\ 0 & 0 & 1 & 0 \\ -\gamma\beta & 0 & 0 & \gamma \end{pmatrix} \begin{pmatrix} \gamma & 0 & 0 & -\gamma\beta' \\ 0 & 1 & 0 & 0 \\ 0 & 0 & 1 & 0 \\ -\gamma\beta' & 0 & 0 & \gamma \end{pmatrix} = \begin{pmatrix} 1 & 0 & 0 & 0 \\ 0 & 1 & 0 & 0 \\ 0 & 0 & 1 & 0 \\ 0 & 0 & 0 & 1 \end{pmatrix}, \tag{II-3}$$

namely $\Lambda^{\mu}_{\sigma} \Lambda'^{\sigma}_{\nu} = \delta^{\mu}_{\nu}$. The Minkowski distance is described by a quadratic about a metric matrix, given by

$$g^{\mu\nu} X_{\mu} X_{\nu} = g_{\mu\nu} X^{\mu} X^{\nu} = (x \ y \ z \ ct) \begin{pmatrix} -1 & 0 & 0 & 0 \\ 0 & -1 & 0 & 0 \\ 0 & 0 & -1 & 0 \\ 0 & 0 & 0 & 1 \end{pmatrix} \begin{pmatrix} x \\ y \\ z \\ ct \end{pmatrix} = (ct)^2 - x^2 - y^2 - z^2, \tag{II-4}$$

where the metric tensors are given by $g_{\mu\nu} = g^{\mu\nu} = diag(-1,-1,-1,+1)$, with $g_{\alpha\beta} = g_{\mu\nu} \Lambda^{\mu}_{\alpha} \Lambda^{\nu}_{\beta}$, or $g = \Lambda^T g \Lambda = \Lambda g \Lambda$, $X^{\mu} = g^{\mu\nu} X_{\nu}$, and $X_{\mu} = g_{\mu\nu} X^{\nu}$.

Based on above, we will derive the Doppler formula for a moving point light source below. Suppose the moving point light source is fixed at the origin of $X'Y'Z'$, namely $\mathbf{x}' = 0$. Thus in the source-rest frame $X'Y'Z'$, required by wave equation, the point-source phase function is given by

$$\Psi' = \omega' t' - |\mathbf{k}'||\mathbf{x}'| = \omega' t' - \left(|\mathbf{k}'|\frac{\mathbf{x}'}{|\mathbf{x}'|}\right) \cdot \mathbf{x}' = \omega' t' - \mathbf{k}'_p \cdot \mathbf{x}' = g'_{\mu\nu} K'^{\mu} X'^{\nu}, \tag{II-5}$$

where, to reflect the constraint between $(\mathbf{x}', ct')$ and $(\mathbf{k}', \omega'/c)$ for a moving point light source, the point-source wave vector is written as $\mathbf{k}'_p = |\mathbf{k}'| \frac{\mathbf{x}'}{|\mathbf{x}'|} = \frac{\omega'}{c} \frac{\mathbf{x}'}{|\mathbf{x}'|}$, the metric tensor is given by $g'_{\mu\nu} = diag(-1,-1,-1,+1)$, and an *assumption* of the Lorentz covariance of $(\mathbf{k}'_p, \omega'/c)$ is *used*. $K'^{\mu} = (\mathbf{k}'_p, \omega'/c)$ and $X'^{\nu} = (\mathbf{x}', ct')$ are given by

$$K'^{\mu} = (\mathbf{k}'_p, \omega'/c) \rightarrow \begin{pmatrix} k'_{px} \\ k'_{py} \\ k'_{pz} \\ \frac{\omega'}{c} \end{pmatrix} = \frac{\omega'}{c} \begin{pmatrix} \frac{x'}{\sqrt{x'^2 + y'^2 + z'^2}} \\ \frac{y'}{\sqrt{x'^2 + y'^2 + z'^2}} \\ \frac{z'}{\sqrt{x'^2 + y'^2 + z'^2}} \\ 1 \end{pmatrix}, \qquad X'^{\nu} = (\mathbf{x}', ct') \rightarrow \begin{pmatrix} x' \\ y' \\ z' \\ ct' \end{pmatrix}. \qquad \text{(II-6)}$$

The Lorentz transformation of $K^{\mu} = (\mathbf{k}_p, \omega/c)$ from $K'^{\mu} = (\mathbf{k}'_p, \omega'/c)$ is given by

$$\begin{pmatrix} k_{px} \\ k_{py} \\ k_{pz} \\ \frac{\omega}{c} \end{pmatrix} = \begin{pmatrix} \gamma & 0 & 0 & -\gamma\beta' \\ 0 & 1 & 0 & 0 \\ 0 & 0 & 1 & 0 \\ -\gamma\beta' & 0 & 0 & \gamma \end{pmatrix} \begin{pmatrix} k'_{px} \\ k'_{py} \\ k'_{pz} \\ \frac{\omega'}{c} \end{pmatrix}. \qquad \text{(II-7)}$$

We have $k'_{px} = \frac{\omega'}{c} \frac{x'}{\sqrt{x'^2 + y'^2 + z'^2}}$ from Eq. (II-6), and from above we have the Lorentz-transformed frequency, given by

$$\frac{\omega}{c} = -\gamma\beta' k'_{px} + \gamma \frac{\omega'}{c} = -\gamma\beta' \left( \frac{\omega'}{c} \frac{x'}{\sqrt{x'^2 + y'^2 + z'^2}} \right) + \gamma \frac{\omega'}{c} = \gamma \frac{\omega'}{c} \left( -\frac{\beta' x'}{\sqrt{x'^2 + y'^2 + z'^2}} + 1 \right). \qquad \text{(II-8)}$$

Re-writing the above equation in a 3D-vector form, we have

$$\frac{\omega}{c} = \frac{\omega'}{c} \gamma \left( 1 - \boldsymbol{\beta}' \cdot \frac{\mathbf{x}'}{|\mathbf{x}'|} \right). \qquad \text{(II-9)}$$

which is exactly the same as Eq. (I-8) given in Attachment-I, with a singularity at the point $\mathbf{x}' = 0$.

Again from Eq. (II-7), we have

$$k_{px} = \gamma k'_{px} - \gamma\beta' \frac{\omega'}{c}, \qquad k_{py} = k'_{py}, \qquad k_{pz} = k'_{pz}. \qquad \text{(II-10)}$$

Re-writing the above equation in a 3D-vector form, we have

$$\mathbf{k}_p = \mathbf{k}'_p + \frac{\gamma - 1}{\beta'^2} (\boldsymbol{\beta}' \cdot \mathbf{k}'_p) \boldsymbol{\beta}' - \gamma \boldsymbol{\beta}' \left( \frac{\omega'}{c} \right), \qquad \text{(II-11)}$$

namely Eq. (I-7) given in Attachment-I.

It is seen from Eq. (II-9) that, when the observer and the point source overlap ($\mathbf{x}' = 0$), observed in the lab frame the frequency is indeterminate.

**Conclusion:** The assumption of the Lorentz covariance of $(\mathbf{k}', \omega'/c)$ for a moving point light source is not physical.

**Attachment-III: Unit wave vector in the lab frame for a moving point light source**

As seen in the derivation of point-source Doppler formula given in Sec. III, the wavefront received at the observation time-space point $(\mathbf{x},t)$ in the lab frame is emitted by the source at the advanced time-space point $(\mathbf{x}_{sa},t_a)$, with $|\mathbf{x}-\mathbf{x}_{sa}|=c(t-t_a)$, which is a Lorentz invariant expression. Thus the unit wave vector at the observation point in the lab frame is given by

$$\hat{\mathbf{n}} = \frac{(\mathbf{x}-\mathbf{x}_{sa})}{|\mathbf{x}-\mathbf{x}_{sa}|}. \tag{III-1}$$

Here we will show that this unit wave is the same as the one given by Eq. (I-11).

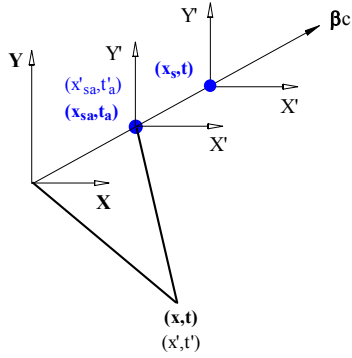

Fig. III-1. Unit wave vector for a moving point source is defined as $\hat{\mathbf{n}} = (\mathbf{x}-\mathbf{x}_{sa})/|\mathbf{x}-\mathbf{x}_{sa}|$. The wavefront emitted by a moving point source at an advanced time-space point $(\mathbf{x}_{sa},t_a)$ reaches the observation point $(\mathbf{x},t)$ when the source reaches $(\mathbf{x}_s,t)$. Observed in the source-rest frame $X'Y'Z'$, $\mathbf{x}'_s = 0$ holds at any time because the point source is fixed at the origin of $X'Y'Z'$, which moves at $\boldsymbol{\beta} = -\boldsymbol{\beta}'$ with respect to the lab frame $XYZ$.

Suppose that the source-rest frame moves at $\boldsymbol{\beta} = -\boldsymbol{\beta}'$ with respect to the lab frame $XYZ$, as shown in Fig. III-1. In the source-rest frame, the corresponding observation time-space point is $(\mathbf{x}',t')$ and the advanced time-space point is $(\mathbf{x}'_{sa}=0,t'_a)$, with $|\mathbf{x}'-\mathbf{x}'_{sa}|=|\mathbf{x}'|=c(t'-t'_a)$, since the source is assumed to be fixed at the origin of $X'Y'Z'$ frame.

With $\mathbf{x}'-\mathbf{x}'_{sa}=\mathbf{x}'$ and $|\mathbf{x}'-\mathbf{x}'_{sa}|=|\mathbf{x}'|=c(t'-t'_a)$ taken into account, we have the Lorentz transformation

$$\mathbf{x}-\mathbf{x}_{sa} = \mathbf{x}' + \frac{\gamma-1}{\beta^2}\mathbf{x}'\cdot\boldsymbol{\beta}'\boldsymbol{\beta}' - \gamma\boldsymbol{\beta}'c(t'-t'_a) = |\mathbf{x}'|\left(\frac{\mathbf{x}'}{|\mathbf{x}'|} + \frac{\gamma-1}{\beta^2}\frac{\mathbf{x}'}{|\mathbf{x}'|}\cdot\boldsymbol{\beta}'\boldsymbol{\beta}' - \gamma\boldsymbol{\beta}'\right). \tag{III-2}$$

Considering $|\mathbf{x}-\mathbf{x}_{sa}|=c(t-t_a)$, from the above Eq. (III-2) we have

$$\hat{\mathbf{n}} = \frac{(\mathbf{x}-\mathbf{x}_{sa})}{|\mathbf{x}-\mathbf{x}_{sa}|} = \frac{(\mathbf{x}-\mathbf{x}_{sa})}{c(t-t_a)} = \frac{|\mathbf{x}'|}{c(t-t_a)}\left(\frac{\mathbf{x}'}{|\mathbf{x}'|} + \frac{\gamma-1}{\beta^2}\frac{\mathbf{x}'}{|\mathbf{x}'|}\cdot\boldsymbol{\beta}'\boldsymbol{\beta}' - \gamma\boldsymbol{\beta}'\right). \tag{III-3}$$

The Lorentz transformation of $c(t-t_a)$ is given by

$$c(t-t_a) = \gamma[c(t'-t'_a) - \boldsymbol{\beta}'\cdot(\mathbf{x}'-\mathbf{x}'_{sa})] = \gamma(|\mathbf{x}'|-\boldsymbol{\beta}'\cdot\mathbf{x}') = |\mathbf{x}'|\gamma\left(1-\boldsymbol{\beta}'\cdot\frac{\mathbf{x}'}{|\mathbf{x}'|}\right), \tag{III-4}$$

where $|\mathbf{x}'-\mathbf{x}'_{sa}|=|\mathbf{x}'|=c(t'-t'_a)$ is employed.

Inserting Eq. (III-4) into Eq. (III-3), we have

$$\hat{\mathbf{n}} = \frac{\dfrac{\mathbf{x}'}{|\mathbf{x}'|} + \dfrac{\gamma-1}{\beta^2}\dfrac{\mathbf{x}'}{|\mathbf{x}'|}\cdot\boldsymbol{\beta}'\boldsymbol{\beta}' - \gamma\boldsymbol{\beta}'}{\gamma\left(1 - \boldsymbol{\beta}'\cdot\dfrac{\mathbf{x}'}{|\mathbf{x}'|}\right)} , \tag{III-5}$$

which are exactly the same as Eq. (I-11) given in Attachment-I.

From Eq. (III-1), we can see that the unit wave vector points from the *advanced* source position to the observation position. At the same place, the unit wave vector changes with time because the source is moving. Observed at different locations at the same time, the unit wave vector is also different, because the retarded times and retarded source positions are different. [Note: For a plane wave, observed in any given inertial frame, the unit wave vector and the frequency are the same everywhere.]

It should be noted that the unit wave vector is indeterminate at the overlapping point ($\mathbf{x}' = 0$); that is because the distance between the source and observation points is zero; zero vector has no determinate direction. However the wave period has a determinate definition given in Sec. III.

The unit-wave-vector transformation Eq. (III-5) also can be obtained from the Lorentz transformation of photon's momentum and energy. As indicated in Sec. III and Appendix C, $(\omega/c)\hat{\mathbf{n}}$ and $(\omega/c)$ cannot constitute a 4-vector and $\hbar$ is not a Lorentz invariant, but the photon's momentum $\mathbf{p} = (\hbar\omega/c)\hat{\mathbf{n}}$ and the energy $E = \hbar\omega$ constitute a 4-vector $P^\mu = (\mathbf{p}, E/c)$. From Lorentz transformation, we have

$$\frac{\hbar\omega}{c}\hat{\mathbf{n}} = \left(\frac{\hbar'\omega'}{c}\hat{\mathbf{n}}'\right) + \frac{\gamma-1}{\beta^2}\boldsymbol{\beta}'\cdot\left(\frac{\hbar'\omega'}{c}\hat{\mathbf{n}}'\right)\boldsymbol{\beta}' - \gamma\boldsymbol{\beta}'\left(\frac{\hbar'\omega'}{c}\right), \tag{III-6}$$

$$\frac{\hbar\omega}{c} = \gamma\left[\left(\frac{\hbar'\omega'}{c}\right) - \boldsymbol{\beta}'\cdot\left(\frac{\hbar'\omega'}{c}\hat{\mathbf{n}}'\right)\right], \tag{III-7}$$

where the transformation between $\omega$ and $\omega'$ is defined by Eq. (4) in Sec. III. From Eq. (III-6)/Eq. (III-7), we have

$$\hat{\mathbf{n}} = \frac{\hat{\mathbf{n}}' + \dfrac{\gamma-1}{\beta^2}(\boldsymbol{\beta}'\cdot\hat{\mathbf{n}}')\boldsymbol{\beta}' - \gamma\boldsymbol{\beta}'}{\gamma(1 - \boldsymbol{\beta}'\cdot\hat{\mathbf{n}}')} , \tag{III-8}$$

where $\hat{\mathbf{n}}' = \mathbf{x}'/|\mathbf{x}'|$ is the unit wave vector in the source-rest frame.

From above, we see that Eq. (III-8) is indeed the same as Eq. (III-5).